\documentclass[12pt]{article}
\usepackage{amsmath, amssymb, enumerate, amsthm}
\usepackage{bm}
\usepackage{graphicx}
\usepackage{booktabs}
\usepackage{siunitx}
\usepackage{hyperref}
\usepackage{natbib}

\newcommand{\ds}{\,{\rm d}s}
\newcommand{\dt}{\,{\rm d}t}

\newcommand{\Var}{{\rm Var}}
\newcommand{\Cov}{{\rm Cov}}

\newcommand{\eqae}{\overset{{\rm a. e.}}{=}}

\newtheorem{Procedure}{Procedure}

\theoremstyle{plain}% Theorem-like structures provided by amsthm.sty

\newtheorem{Remark}{Remark}
\newtheorem{Lemma}{Lemma}
\newtheorem{Corollary}{Corollary}

\newcommand{\blind}{1}

\begin{document}

\def\spacingset#1{\renewcommand{\baselinestretch}%
{#1}\small\normalsize} \spacingset{1}

%%%%%%%%%%%%%%%%%%%%%%%%%%%%%%%%%%%%%%%%%%%%%%%%%%%%%%%%%%%%%%%%%%%%%%%%%%%%%%

\if1\blind
{
  \title{\bf Localized Functional Principal Component Analysis Based on Covariance Structure}
  \author{Maria Laura Battagliola \\[0.1cm]
  Department of Statistics, Instituto Tecnológico Autónomo de México,\\ Mexico City, Mexico \\[0.6cm]
  Jan O. Bauer \\[0.1cm]
    Department of Econometrics and Data Science, Vrije Universiteit Amsterdam, \\
    Amsterdam, Netherlands\\
    Tinbergen Institute, Amsterdam, Netherlands}
  \maketitle
} \fi

\if0\blind
{
  \bigskip
  \bigskip
  \bigskip
  \begin{center}
    {\LARGE\bf Localized Functional Principal Component Analysis Based on Covariance Structure}
\end{center}
  \medskip
} \fi

\bigskip
\begin{abstract}
\noindent
Functional principal component analysis (FPCA) is a widely used technique in functional data analysis for identifying the primary sources of variation in a sample of random curves. The eigenfunctions obtained from standard FPCA typically have non-zero support across the entire domain. In applications, however, it is often desirable to analyze eigenfunctions that are non-zero only on specific portions of the original domain—and exhibit zero regions when little is contributed to a specific direction of variability—allowing for easier interpretability. Our method identifies sparse characteristics of the underlying stochastic process and derives localized eigenfunctions by mirroring these characteristics without explicitly enforcing sparsity. Specifically, we decompose the stochastic process into uncorrelated sub-processes, each supported on disjoint intervals. Applying FPCA to these sub-processes yields localized eigenfunctions that are naturally orthogonal. In contrast, approaches that enforce localization through penalization must additionally impose orthogonality. Moreover, these approaches can suffer from over-regularization, resulting in eigenfunctions and eigenvalues that deviate from the inherent structure of their population counterparts, potentially misrepresenting data characteristics. Our approach avoids these issues by preserving the inherent structure of the data. Moreover, since the sub-processes have disjoint supports, the eigenvalues associated to the localized eigenfunctions allow for assessing the importance of each sub-processes in terms of its contribution to the total explained variance. We illustrate the effectiveness of our method through simulations and real data applications. Supplementary material for this article is available online.
\end{abstract}

\noindent%
{\it Keywords:}  Domain selection; Functional data analysis; Functional principal component analysis; Null region; Sparse principal component analysis 
\vfill

\newpage
\spacingset{1.9} % DON'T change the spacing!

\section{Introduction}
Nowadays, large volumes of data are readily available like never before. Due to recordings on high resolution, many of the observed phenomena possess a latent continuous nature. This holds true for several processes, from environmental sciences to biomedical imaging. Functional data analysis (FDA) is the branch of statistics that studies samples of observations that can be thought as discrete recordings of an underlying smooth stochastic process $\mathcal{X}(\cdot)$ that is unknown in practice. For a comprehensive review of the topic, we refer to \citet{ramsay2005functional}, \citet{wang2016functional}, or \citet{crainiceanu2024functional}.

Principal component analysis (PCA) has long been a key tool for dimensionality reduction in multivariate data analysis, particularly when data dimensionality is very high. By approximating the data with the first few principal components, PCA aims to visualize important structures and extract meaningful patterns. Similarly to the multivariate case, it is possible to perform functional principal component analysis (FPCA) for FDA via the spectral decomposition of the integral covariance operator (see, e.g., \citet{rice1991}, \citet{Yao05}, \citet{hall2006}, among others). The spectral decomposition consists of eigenvalues and eigenfunctions which allow for checking the main modes of variation of the functional data.

One drawback of FPCA is that its eigenfunctions typically have support coinciding with the entire domain, which complicates interpretation due to the lack of sparsity. To overcome this, methods have been proposed that directly estimate compactly supported eigenfunctions, thereby improving interpretability \citep{chen2015localized, LWC16, nie2020sparse}. These approaches enforce compact supports via regularization and—because sample curves are often observed on a dense, regular grid—employ multivariate penalties such as SCAD \citep{FL01} generalized to the functional data setting.

However, if the eigenfunctions are over-regularized, they can deviate greatly from the population eigenfunctions of the underlying stochastic process $\mathcal{X}(\cdot)$. This leads to overestimation or underestimation of the corresponding eigenvalues and, consequently, of the explained variance. Moreover, localized eigenfunctions obtained by regularization are not necessarily orthogonal, which complicates the calculation of the explained variance, as the contributions of the individual components must be disentangled. As in cross-sectional settings (see, e.g., \citet{ZHT06}, \citet{SH08}, or \citet{WLL09} among others) this is also a concern for functional data \citep{nie2020sparse}.

In this work, we therefore propose a {\em localized} FPCA (L-FPCA), whose eigenfunctions reflect the inherent structure of the underlying stochastic process. Specifically, we decompose $\mathcal{X}(\cdot)$ into uncorrelated sub-processes, meaning that the autocovariance function is compactly supported and exhibits uncorrelated {\em sub-intervals}.  Detection of such sub-intervals at which sampled functions show similarities or dissimilarities has been of raising interest \citep{pini2016interval, pini2017interval,blanquero2020selection,blanquero2023optimal,beyaztas2022functional}. Our proposed methodology identifies leading domains of the stochastic process which exhibit most variability. Further, the eigenfunctions of $\mathcal{X}(\cdot)$ are inherently sparse under such a decomposition into uncorrelated sub-processes. Our proposed L-FPCA naturally aligns with this structure, enhancing interpretability while preserving the representation of the underlying stochastic process. Additionally, the resulting localized eigenfunctions are orthogonal by construction. This means they do not share information about the explained variance, and their individual contribution is thus given by their eigenvalues.

Since the sample curves are recorded on a grid, the observed sample autocovariance—hereafter referred to as the {\em covariance} function throughout this work—takes the form of a covariance matrix.
 Several efforts have been made to estimate smooth covariance functions from functional observations (see, e.g., \citet{xiao2016fast}, \citet{xiao2018fast}, \citet{delaigle2021estimating}, or \citet{wang2022low}, among others). By proposing localized eigenfunctions that capture the inherent structure of the underlying stochastic process, we focus on covariance functions in which correlated grid points occur only within individual sub-processes. We will show that this property occurs under block diagonal covariance matrix structures. In multivariate settings, testing for block diagonal covariance matrices is of interest for many real-world application such as Gaussian graphical models or estimation of covariance matrices. However, identifying these block diagonal structures can be challenging in high-dimensional, low-sample-size settings, or in general when the number of observations is small which is often the case in FDA. Therefore, several methods have been suggested for such task (see, e.g., \citet{DG18},  \citet{SWB24}, \citet{BA24}, and references therein).

\begin{figure}[!htbp]
\center
\includegraphics[width=0.7\textwidth]{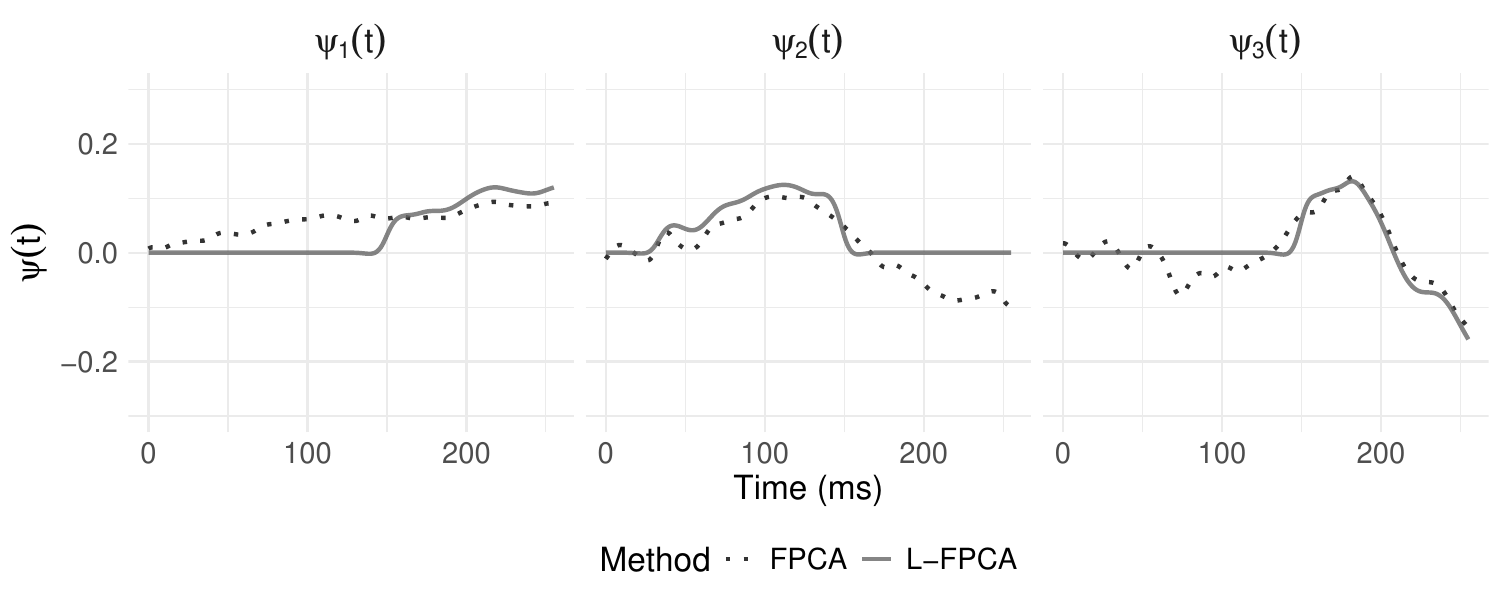}
  \caption{Eigenfunctions (\textit{dotted black}) and localized eigenfunctions (\textit{solid gray}) of an electroencephalogram dataset ranging from $0$ to $256$ Hz (3.9-ms epoch).}
  \label{fig:Introduction}
\end{figure}

An illustration of the methodology proposed in this work is provided in Figure~\ref{fig:Introduction}, which shows the first three eigenfunctions of an electroencephalogram (EEG) dataset containing $N_A = 77$ signal records of alcoholic patients ranging from $0$ to $256$ Hz (3.9-ms epoch). Eigenfunctions from traditional FPCA are represented with dotted black curves, and the localized eigenfunctions computed using L-FPCA are solid gray lines, respectively. The localized eigenfunctions exhibit behavior similar to that of the traditional ones on compact supports, and equal zero elsewhere. For example, the first and third localized eigenfunctions are approximately supported from the 150th millisecond onward, while the second localized eigenfunction is supported in the first part of the time domain. Computational details, data information, and further elaborations for this example are discussed in Section~\ref{sec:RealData}.

This work is organized as follows. Section~\ref{sec:Preliminaries} expands on the framework of our suggested method. Then, in Section~\ref{sec:SFPCA}, we introduce the concept of L-FPCA on an underlying stochastic process. A simulation study is presented in Section~\ref{sec:SimulationStudy} along with some practical considerations. We provide two real data applications of the proposed methodology in Section~\ref{sec:RealData}, and we discuss our findings and further extensions in Section~\ref{sec:Discussion}.

\section{Preliminaries}
\label{sec:Preliminaries}

Let $\mathcal{X}: \mathcal{T}  \mapsto \mathbb{R}$ be a mean zero continuous stochastic process over a compact interval $\mathcal{T} =  [t_0 , t_K]\subset \mathbb{R}$ with covariance function $\Gamma(s,t) = \Cov(\mathcal{X}(s),\mathcal{X}(t))$. Moreover, we assume that the process is square-integrable, i.e., $\mathcal{X}(\cdot) \in L^2(\mathcal{T})$.
We define the covariance operator as
\begin{equation}\label{eq:CovarianceOperator}
    (\mathcal{G}\psi)(t) = \int_{\mathcal{T}} \Gamma(s,t) \psi(s) \ds \,,
\end{equation}
for $t \in \mathcal{T}$. The stochastic processes $\mathcal{X}(\cdot)$ and $\mathcal{Y}(\cdot)$ are said to be equal almost everywhere, denoted as $\mathcal{X}(\cdot) \eqae \mathcal{Y}(\cdot)$, if they coincide except for a non-measurable set. For the purpose of this work, we partition $\mathcal{X}(\cdot)$ into $b$ compactly supported sub-processes, such that 
\begin{equation}\label{eq:sub-processes}
    \mathcal{X}(t) \overset{{\rm a. e.}}{=} \sum\limits_{k=1}^K \mathcal{X}_k(t) \,, \;  \mathcal{X}_k(t) \overset{{\rm a. e.}}{=}  \begin{cases}
        \mathcal{X}(t) & ,\,  t \in \mathcal{T}_k \\
        0              & ,\,  t \not\in \mathcal{T}_k
    \end{cases} \;,\, k \in \{1, \ldots, K\} \,,
\end{equation}
where $\mathcal{T}_k = [t_{k-1}, t_k]$ with $t_{k-1} \leq t_k$ thus $\mathcal{T}_k \cap \mathcal{T}_{k+1} = t_{k}$ and $\bigcup_{k=1}^K \mathcal{T}_k = \mathcal{T}$. Consequently, all $\mathcal{X}_k: \mathcal{T} \mapsto \mathbb{R}$ are square-integrable processes over the same domain as $\mathcal{X}(\cdot)$. The partition of $\mathcal{X}(\cdot)$ into the sub-processes holds almost everywhere except for the non-measurable intersections $\mathcal{T}_k \cap \mathcal{T}_{k+1} = t_{k}$. This construction ensures that all $\mathcal{T}_k$ are compact. In applications, the processes are observed over discrete grids making the observed intervals disjoint. Moreover, $t_{k-1} \leq t_k$ assumes that the intervals are arranged in ascending order. This is made for notational convenience and imposes no restriction on the proposed methodology (a discussion is given in B~Online Appendix in the supplementary material). Further, we assume that 
\begin{equation}\label{eq:NoAutocorrelation}
    \Gamma(s,t) \eqae 0\,; \;\; s\in\mathcal{T}_k,\, t\in\mathcal{T}_{k'} ,\, k \neq k' \,,
\end{equation}
namely that the distinct sub-processes have autocorrelation of zero almost everywhere.

According to Mercer's theorem, the covariance function $\Gamma(s,t)$ can be represented as
\begin{equation*}
    \Gamma(s,t) = \sum\limits_{l=1}^\infty \lambda_l \psi_l(s) \psi_l(t) \,, 
\end{equation*}
where $\{\psi_l(\cdot)\}_{l=1}^\infty$ are eigenfunctions with corresponding eigenvalues $\{\lambda_i\}_{i=1}^\infty$ of the covariance function, ordered such that $\lambda_1 \geq \lambda_2 \geq \cdots > 0$. The assumption of positive eigenvalues is not restrictive in practical applications, as we generally work with truncated, finite-dimensional representations of stochastic processes. This assumption is further discussed in Section~\ref{sec:SFPCA}. The eigenfunctions are orthonormal, i.e.
\begin{equation*}
    \int_\mathcal{T} \psi_l^2(t) \dt = 1 \;\text{ and }\; \int_\mathcal{T} \psi_l(t) \psi_{l'}(t)  \dt = 0 \, , \, l \neq l' \,.
\end{equation*}
Owing to these properties, the eigenfunctions $\{\psi_i(\cdot)\}_{i=1}^\infty$ constitute a basis of $L^2(\mathcal{T})$. Therefore, $\mathcal{X}(\cdot)$ can be represented as
\begin{equation}
\label{eq:KL}
    \mathcal{X}(t) = \sum\limits_{l=1}^\infty \xi_l \psi_l(t) \,,
\end{equation}
where the $\{\xi_l\}_{l=1}^\infty$, with $\Var(\xi_l) = \lambda_l$, are the FPC scores corresponding to the eigenfunctions. Representation \eqref{eq:KL} is called the Karhunen-Loève (KL) expansion of the process $\mathcal{X}(\cdot)$.

For $k\in\{1, \ldots, K\}$, each sub-process $\mathcal{X}_k(\cdot)$ has covariance function $\Gamma_k(s,t)  = \Cov (\mathcal{X}_k(s),\mathcal{X}_k(t))$. Due to \eqref{eq:NoAutocorrelation}, it holds for the covariance functions that $ \Gamma_k(s,t) = 0$ if $s\not \in \mathcal{T}_k$ and/or $t\not \in \mathcal{T}_k$, and that $\Gamma_k(s,t) \eqae \Gamma(s,t) $ for $(s,t) \in \mathcal{T}_k \times \mathcal{T}_k$. Furthermore, each $\mathcal{X}_k(\cdot)$ has its own KL expansion:
\begin{equation}\label{eq:KLexpensionsub-processes}
    \mathcal{X}_k(t) = \sum\limits_{j=1}^\infty \xi_{k,j} \psi_{k,j}(t) \,.
\end{equation}
Here, $\{ \psi_{k,j}(\cdot) \}_{j=1}^\infty$ and $\{ \xi_{k,j} \}_{j=1}^\infty$ are the eigenfunctions and scores of $\mathcal{X}_k(\cdot)$ respectively. Consequently, the same propositions that apply to the covariance function $\Gamma(s,t)$ and the KL expansion of the stochastic process $\mathcal{X}(\cdot)$ also extend to the covariance function $\Gamma_k(s,t)$ and the KL expansions of the sub-processes.

\section{Localized Functional Principal Components Analysis}
\label{sec:SFPCA}

In this section, we discuss the methodology for L-FPCA. We demonstrate that if the underlying process $\mathcal{X}(\cdot)$ has eigenfunctions that exhibit sparse charateristics, so do the eigenfunctions of its sub-processes. This can be utilized to compute localized eigenfunctions for $\mathcal{X}(\cdot)$ via the eigenfunctions of its sub-processes.

First, note that $\mathcal{X}(\cdot)$ can be expressed through the KL expansion of its sub-processes.

\begin{Remark}\label{r:RepAsKLSub}
Let $\mathcal{X}(\cdot)$ be partitioned into sub-processes $\mathcal{X}_k(\cdot)$ with $k \in\{1 , \ldots, K\}$ as in (\ref{eq:sub-processes}), where each sub-process exhibits its KL expansion as in (\ref{eq:KLexpensionsub-processes}). For each sub-interval $\mathcal{T}_k$, it holds that
\begin{equation*}
    \mathcal{X}(t) \eqae \sum\limits_{j=1}^\infty \xi_{k,j} \psi_{k,j}(t) \text{ for } t \in \mathcal{T}_{k} \,.
\end{equation*}
\end{Remark}

Next, we analyse the structure of the eigenfunctions associated with the sub-processes. We note that these eigenfunctions exhibit sparse structures, mirroring the block-diagonal characteristics introduced in \eqref{eq:NoAutocorrelation}.
%zero correlation between the sub-processes as established in (\ref{eq:NoAutocorrelation}). 

\begin{Lemma}\label{l:EigenfunctionsOfSubMirrorCov}
Let $\Gamma(s,t)$ be the covariance function of $\mathcal{X}(\cdot)$. It holds that $\Gamma(s,t)  \eqae 0 $ for $s \not\in \mathcal{T}_{k}$ and $t\in\mathcal{T}_{k}$ if and only if the eigenfunctions $\{\psi_{k,j}(\cdot)\}_{j=1}^\infty$ of sub-process $\mathcal{X}_k(\cdot)$ are compactly supported on $\mathcal{T}_k$, namely
\begin{equation}\label{eq:SparseSubEigenfunctions}
\psi_{k,j}(t) =
    \begin{cases} 
    \psi_{k,j}(t)   ,\, & t \in \mathcal{T}_{k} \\
    0               ,\, & t \notin \mathcal{T}_{k}
    \end{cases} \,,
\end{equation}
for $k \in\{1 , \ldots, K \}$.
\end{Lemma}
In particular, Lemma~\ref{l:EigenfunctionsOfSubMirrorCov} demonstrates that the sparse structure of the covariance function allows us to effectively derive localized eigenfunctions for the sub-processes. Such eigenfunctions coincide with those of $\mathcal{X}(\cdot)$ on the support of each sub-process, as we conclude in the following result. Same holds for the eigenvalues.

%Consequently, by identifying the structure of the covariance function, we effectively derive sparse eigenfunctions for the sub-processes thus sparse functional principal components. Moreover, these eigenfunctions of the sub-processes coincide with those of $\mathcal{X}(t)$, thereby providing sparse eigenfunctions for the entire stochastic process.

\begin{Lemma}\label{l:EigenfunctionsAlign}
Let $\mathcal{X}(\cdot)$ be partitioned into sub-processes $\mathcal{X}_k(\cdot)$ as in (\ref{eq:sub-processes}), where each sub-process is represented by its own KL expansion as in (\ref{eq:KLexpensionsub-processes}). For the eigenfunctions and eigenvalues of each sub-process $\mathcal{X}_k(\cdot)$, there are eigenfunctions and eigenvalues of $\mathcal{X}(\cdot)$ such that 
\begin{equation*}
    \{ \psi_{k,j}(t) , \lambda_{k,j}\}_{j=1}^\infty = \{ \psi_{l}(t) , \lambda_{l} \}_{l \subset \{1, 2, \ldots\}} \,,
\end{equation*}
for $t \in \mathcal{T}_k$.
\end{Lemma}
We know that on the domain $\mathcal{T}_k$, the eigenfunctions and eigenvalues of $\mathcal{X}(\cdot)$ coincide with the eigenfunctions and eigenvalues of the sub-process $\mathcal{X}_k(\cdot)$. These eigenfunctions and eigenvalues of the process $\mathcal{X}(\cdot)$ are indexed by $l \subset \{1, 2, \ldots\}$.
Consequently, the eigenfunctions of $\mathcal{X}(\cdot)$ exhibit the same sparse structure as those of the sub-processes.

\begin{Corollary}\label{col:EigenfunctionsOfXMirrorCov}
Let $t\in\mathcal{T}_{k}$, $k \in\{1 , \ldots, K \}$, and let the covariance function of $\mathcal{X}(\cdot)$ be $\Gamma(s,t)$. It holds that $\Gamma(s,t) \eqae 0$ for $s \not\in \mathcal{T}_{k}$ if and only if all eigenfunctions $\{\psi_l(\cdot)\}_{l=1}^\infty$ of $\mathcal{X}(\cdot)$ exhibit the structure
\begin{equation*}
\psi_l(t) =
    \begin{cases} 
    \psi_l(t)   & ,\,  t \in \mathcal{T}_{k} \\
    0           & ,\, t \notin \mathcal{T}_{k}
    \end{cases} \,,
\end{equation*}
namely they are compactly supported on a single interval $\mathcal{T}_k$.
\end{Corollary}

Therefore, the eigenfunctions of the stochastic process $\mathcal{X}(\cdot)$ exhibit a sparse structure as they are inherently compactly supported. Building on this result, we propose an approach for calculating localized functional principal components by identifying the supports of the sub-processes through the covariance function of $\mathcal{X}(\cdot)$.

Now, we elaborate on this approach. First of all, the processes we have been considering so far are infinite dimensional. Therefore, to apply the developed theory in practice, it is necessary to construct a finite-dimensional approximation. This is typically achieved by truncating the KL representation of each process to its first $L$ eigenfunctions:
\begin{equation*}
\widehat{\mathcal{X}}(t) = \sum_{l=1}^L \xi_{l} \psi_l(t) \,,
\end{equation*}
for $t \in \mathcal{T}$. Here, $L$ is usually chosen based on the percentage of variance explained (PVE), namely by
$$L = \text{min} \left\{\tilde{L}\in \{1, 2, \ldots \}: \frac{\sum_{l=1}^{\tilde{L}} \lambda_l}{\sum_{l=1}^{L^{max}} \lambda_l} \geq {\rm PVE} \right\} \,. $$
Generally, PVE is selected to reach a high percentage, such as 90\% or more. Since eigenvalues equal to zero do not contribute to the explained variance, our assumption of positive eigenvalues is justified.
%This supports our assumption of positive eigenvalues, as eigenvalues equal to zero do not contribute to the explained variance. Therefore, only positive eigenalues remain. 
Moreover, the number of eigenfunctions $J_k$ for each sub-process $\mathcal{X}_k(\cdot)$ can either be determined through a similar procedure or set equal to $L$. The latter approach results in more eigenfunctions but offers the advantage of simplicity. In the following, we outline the detailed procedure for the method we propose.

\begin{Procedure}\label{p:1}
    Computation of localized functional principal components for a stochastic process $\mathcal{X}(\cdot)$ with mean zero.

\begin{enumerate}
    \item\label{StepOneP1} Identify the sparse structure of $\Gamma(s,t)$, thus the supports $\mathcal{T}_1 , \ldots, \mathcal{T}_K$, to effectively decompose $\mathcal{X}(\cdot)$ into its sub-processes $\mathcal{X}_1(\cdot) , \ldots, \mathcal{X}_K(\cdot)$.

    \item Calculate the eigenfunctions $\{\psi_{k,j}(\cdot)\}_{j=1}^{J_k}$ and eigenvalues $\{\lambda_{k,j}\}_{j=1}^{J_k}$ for each sub-process $\mathcal{X}_k(\cdot)$. These eigenfunctions are compactly supported on $\mathcal{T}_k$ and coincide with the eigenfunctions $\{ \psi_{i}(\cdot) \}_{i=1}^\infty$ of process $\mathcal{X}(\cdot)$

    \item\label{StepThreeP1} Order all eigenvalues for all sub-processes by their magnitude, and select the $M \leq L$ largest, which align with the eigenvalues $\lambda_1 , \ldots, \lambda_M$ of $\mathcal{X}(\cdot)$ as shown in Lemma~\ref{l:EigenfunctionsAlign}.
\end{enumerate}
\textbf{Output}: Localized functional principal components $\{ \psi_{l}(\cdot) \}_{l=1}^M$.
\end{Procedure}

%The output consists of the $N$ eigenfunctions, which are sparse by construction, together with the corresponding $N$ eigenvalues. The proposed methodology does not impose sparseness on the eigenfunctions. Instead, it uncovers their sparsity when the underlying stochastic process exhibits a sparse structure. Furthermore, since the eigenvalues of the sub-processes equal the eigenvalues of $\mathcal{X}(t)$, they can be used to calculate the overall explained variance. In practice, however, the zero intervals of the covariance matrix are masked by sample noise. Still, the eigenvalues of the sub-processes can be used as an approximation of the explained variance, as they approximate the eigenvalues of the underlying stochastic process.

The proposed methodology yields $M$ localized eigenfunctions along with their corresponding eigenvalues. The selection of $M$ depends on the specific objectives of the application and can be determined using standard approaches, such as cross-validation or by assessing the percentage of variance explained
\begin{equation}\label{eq:PropVarExpl}
     \sum_{l=1}^{M} \lambda_{l} \Big/ \sum_{l=1}^{L} \lambda_{l}   \,.
\end{equation}
While the method does not explicitly enforce sparsity on the eigenfunctions, it reveals sparsity when the underlying stochastic process exhibits a sparse structure. This results in eigenfunctions that are localized by construction, and are additionally orthonormal

\begin{Corollary}\label{col:orthonormal}
    The localized eigenfunctions $\{ \psi_{l}(\cdot) \}_{l=1}^M$ computed in Procedure~\ref{p:1} are orthonormal.
\end{Corollary}

Since the localized eigenfunctions of the process $\mathcal{X}(\cdot)$ are orthonormal, their eigenvalues can be used to compute the explained variance as in equation \eqref{eq:PropVarExpl}. Furthermore, because the eigenvalues of the sub-processes coincide with those of the full process $\mathcal{X}(\cdot)$, we can effectively use the eigenvalues of the sub-processes instead.

As established in Corollary~\ref{col:EigenfunctionsOfXMirrorCov}, the structure of the covariance function allows the derivation of localized eigenfunctions, a property exploited in Step~\ref{StepOneP1} of Procedure~\ref{p:1}. In practice however, the block structure of the covariance function $\Gamma(s,t)$ is perturbed by noise, since we only observe its empirical counterpart. As a result, the supports $\mathcal{T}_1 , \ldots, \mathcal{T}_b$ of the localized eigenfunctions are masked and detecting these is not trivial. To address this challenge, \citet{BA24} recently proposed a method to identify the underlying block diagonal covariance structure using singular vectors (BD-SVD). This methodology forms the foundation in Procedure~\ref{p:1} for obtaining localized functional principal components, and is recapped in greater detail in the supplementary material.

 \section{Simulation Study}
 \label{sec:SimulationStudy}

 L-FPCA computes localized eigenfunctions that mirror the inherent structure of the underlying stochastic process. Consequently, the methodology also identifies the different supports of the sub-processes alongside the explained variance associated with each support. In this section we carry out a simulation study to asses the robustness of our methodology with respect to these two properties.

\subsection{Study Design}
\label{sec:generated_data_sim}

In this simulation study, we consider a zero-mean stochastic process $\mathcal{X}(\cdot) \in L^2(\mathcal{T})$, with $\mathcal{T} = [0,1]$, with an underlying structure $\mathcal{X}(\cdot) \eqae \mathcal{X}_1(\cdot) + \mathcal{X}_2(\cdot) +  \mathcal{X}_3(\cdot)$. The three sub-processes have support $\mathcal{T}_1 = [0,0.3]$, $\mathcal{T}_2 = [0.3,0.6]$, and $\mathcal{T}_3 = [0.6,1]$, respectively. In order to simulate $N$  realizations of $\mathcal{X}(\cdot)$, we use a truncated KL representation
\begin{equation}
\label{eq:KL_sim}
    X_n(s) =  \sum_{k=1}^8 \sqrt{\lambda_k} \xi_{nk} \psi_{k}(s),
\end{equation}
with eigenvalues $\bm{\lambda}=(\lambda_1,\dots, \lambda_8)$, scores $\xi_{nk} \stackrel{iid}{\sim} \mathcal{N}(0, \sigma^2)$ and eigenfunctions $\{ \psi_k(\cdot)\}_{k=1}^K$. We pick $\bm{\lambda} = (6^2, 4^2, 2^2, 0.5^2, 0.25^2, 0.2^2, 0.15^2, 0.1^2)$ and $\sigma^2 = 1$. We consider a scenario (\textit{design A}), where each sub-process is assigned to one eigenfunction, and a more complex scenario (\textit{design B}), where some $\mathcal{X}_1(\cdot)$ is assigned to two eigenfunctions. In design A, the eigenfunctions are the orthonormalized counterparts of
$$\widetilde{\psi}_1(s)= B_1(s), \quad \widetilde{\psi}_2(s) = B_5(s), \quad \widetilde{\psi}_3(s) = B_9(s),$$
and 
\begin{equation}
\label{eq:Fourier_basis}
    \widetilde{\psi}_j(s) = \begin{cases}
        \sqrt{2}\sin(j \, \pi \, s), \quad j\in \{4,6,8 \},\\
         \sqrt{2}\cos((j+1) \, \pi \, s), \quad j \in \{5,7 \},
\end{cases}
\end{equation}
for $j=4,\dots,8$. We denote by $B_1(\cdot), B_5(\cdot), B_9(\cdot)$ the cubic B-splines defined on the whole domain $\mathcal{T}$ with $10$ total knots, and the remainder  \eqref{eq:Fourier_basis} are Fourier basis functions. We choose the first three eigenfunctions to be B-splines since they are naturally compactly-supported.
This setting is inspired by the simulation study of \citet{chen2015localized}. In our simulation study, $\widetilde{\psi}_1(s) = \widetilde{\psi}_{1,1}(s)$ is a compactly supported eigenfunction of the first sub-process $\mathcal{X}_1(\cdot)$, and similarly we have compactly supported eigenfunctions $\widetilde{\psi}_2(s) = \widetilde{\psi}_{2,1}(s)$ and $\widetilde{\psi}_3(s) = \widetilde{\psi}_{3,1}(s)$ of $\mathcal{X}_2(\cdot)$ and $\mathcal{X}_3(\cdot)$, respectively.

% For our other scenario, we pick $K=11$ eigenfunctions, each sub-process is assigned two eigenfunctions. In particualr, we consider cubic B-splines with $6$ total knots defined on each sub-interval, and we orthonormalize them such that they are valid eigenfunctions for each sub-process $\mathcal{X}_1(\cdot), \mathcal{X}_2(\cdot), \mathcal{X}_3(\cdot)$. Then, we pad them with $0$ outside the corresponding domains. We call $B_{f}^g(\cdot)$ the $f$-th orthonormalized B-spline having support sub-domain $\mathcal{T}_g$. Then, the eigenfunctions for process $\mathcal{X}(\cdot)$ are the orthonormalized version of
% \begin{align*}
%     & \widetilde{\psi}_1(s)= B_2^1(s), \quad \widetilde{\psi}_2(s) = B_4^1(s),\\
%      & \widetilde{\psi}_3(s)= B_2^2(s), \quad \widetilde{\psi}_4(s) = B_4^2(s),\\
%      & \widetilde{\psi}_5(s)= B_2^3(s), \quad \widetilde{\psi}_6(s) = B_4^3(s),\\
% \end{align*}
% and

% \begin{align*}
%     \widetilde{\psi}_j(s) = \begin{cases}
%         \sqrt{2}\sin((j-3) \, \pi \, s), \quad j\in \{7,9,11 \},\\
%          \sqrt{2}\cos((j+2) \, \pi \, s), \quad j \in \{8,10 \},
% \end{cases}
% \end{align*}
% for $j=7,\dots,11$. In this scenario, we assign eigenvalues $\lambda = (0.7\cdot6^2, 0.3\cdot6^2, 0.7\cdot4^2, 0.3\cdot4^2, 0.7\cdot2^2, 0.3\cdot2^2, 0.5^2, 0.25^2, 0.2^2, 0.15^2, 0.1^2)$, namely we assign the same variability to each sub-process as before by partitioning the eigenvalues used to simulate the simple scenario in two.

In design B, we construct two compactly supported eigenfunctions on $\mathcal{T}_1$, rather than one. Specifically, we consider the second and fourth cubic B-splines with six total knots defined on the first sub-interval.
These splines are padded with zeroes outside $\mathcal{T}_1$. We denote by $B_{2}^1(\cdot)$ and $B_{4}^1(\cdot)$ the $f$-th orthonormalized B-spline with support on $\mathcal{T}_1$. The procedure then follows design A, except that the first two not-orthonormalized eigenfunctions of $\mathcal{X}_1(\cdot)$ are given by
$$\widetilde{\psi}_{1,1}(s)= B_{2}^1(s), \quad \widetilde{\psi}_{1,2}(s)= B_{4}^1(s) \,,$$
with corresponding eigenvalues $\lambda_{1,1} = 0.7\cdot\lambda_1$ and $\lambda_{1,2} = 0.3\cdot\lambda_1$.
% \begin{align*}
%     & \widetilde{\psi}_1(s)= B_2^1(s), \quad \widetilde{\psi}_2(s) = B_4^1(s),\\
%      & \widetilde{\psi}_3(s)= B_2^2(s), \quad \widetilde{\psi}_4(s) = B_4^2(s),\\
%      & \widetilde{\psi}_5(s)= B_2^3(s), \quad \widetilde{\psi}_6(s) = B_4^3(s),\\
% \end{align*}
% and

% \begin{align*}
%     \widetilde{\psi}_1(s) &= B_1(s) \\
%     \widetilde{\psi}_2(s) &= B_5(s) \\
%     \widetilde{\psi}_3(s) &= B_9(s) \\
%     \widetilde{\psi}_j(s) &= \begin{cases}
%         \sqrt{2}\sin(j \, \pi \, s), \quad j\in \{4, 6, 8 \}\\
%          \sqrt{2}\cos((j+1) \, \pi \, s), \quad j\in \{5, 7\}
%     \end{cases}
% \end{align*}
% whose eigenfunctions $\psi_1(\cdot), \psi_2(\cdot),\psi_3(\cdot)$ are compactly supported on three distinct sub-intervals $\mathcal{T}_1 = [0,2]$, $\mathcal{T}_2 = [2,6]$, and $\mathcal{T}_3 = [6,10]$, respectively. Consequently, we consider $\mathcal{X}(\cdot) \eqae \mathcal{X}_1(\cdot) + \mathcal{X}_2(\cdot) +  \mathcal{X}_3(\cdot)  $ with $K = 3$ as in \eqref{eq:sub-processes}. 

Lastly, since functional data is recorded on a grid with sample noise in practice, we build
\begin{equation}
\label{eq:noisy_data}
    W_{np} = X_n(s_p) + \epsilon_{np} \,.
\end{equation}
The functions in \eqref{eq:KL_sim} are evaluated on a dense grid $\mathcal{S}=(s_1,\dots, s_P)$, with $P = 1001$ equidistant points such that $0=s_1 < \dots <s_P = 1$. We denote the sample grid of the compact supports $\mathcal{T}_1$, $\mathcal{T}_2$, and $\mathcal{T}_3$ as $\mathcal{S}_1$, $\mathcal{S}_2$, and $\mathcal{S}_3$ respectively. Then, we generate data comprising $N\in \{250, 500, 1000\}$ curves $R=500$ times realized on $\mathcal{S}$. We pick $M=10$ eigenfunctions
in all our scenarios, so that we can compare more effectively the basis functions with the detected eigenfunctions.

%We also compare our method with two approaches: a naive approach based on a simple thresholding procedure on the computed eigenfunctions, and the one proposed by \cite{nie2020sparse}. We have shown in Corollary~\ref{col:EigenfunctionsOfXMirrorCov} that the eigenfunctions of a stochastic process that consists of uncorrelated sub-processes mirror the support of these sub-processes. Consequently, a naive approach is to identify these supports by thresholding the observed eigenfunctions, as suggested by \citet{BD21} for eigenvectors in a multivariate data setting. Specifically, we select an arbitrary threshold value and identify the intervals where the absolute value of the estimated eigenfunctions exceeds this threshold. The thresholds considered are \(10^{-3}\) and \(10^{-7}\), respectively. Note that this approach may yield multiple noncontiguous intervals; therefore, we define the final selected interval as the widest interval among those identified. We apply the fast covariance estimation (FACE-FPCA) with a sandwich smoother \citep{xiao2013, xiao2016fast} \citep{xiao2013, xiao2018fast}. We use the function \texttt{fpca.face} implemented in the \texttt{refund} package \citep{refund} in \texttt{R}. %Finally, we fix the number of FPCs to $L=6$ in both procedures as in our application in Section \ref{sec:RealData}.
We compare our method with two other approaches:
\begin{enumerate}
    \item[] \textit{FPCA($\tau$)}: This approach is based on a simple thresholding procedure on the computed eigenfunctions. Corollary~\ref{col:EigenfunctionsOfXMirrorCov} establishes that the eigenfunctions of a stochastic process mirror the supports of its covariance function. Thus, similar to the method of \citet{BD21} for cross-sectional data, an ad-hoc approach to recover these supports is to identify all regions where the absolute value of the estimated eigenfunctions is larger than a certain threshold. In this simulation study, we selected $\tau = 10^{-3}$ and $ \tau=  10^{-5}$, and implementation details are provided in the supplementary material.

\item[] \textit{SFPCA}: \citet{nie2020sparse} derived a method to compute localized (sparse) eigenfunctions using a regression-type penalization approach. The penalization includes an elastic-net regularization alongside a roughness part for smoothing. The authors show that SFPCA outperforms the existing methods of \citet{chen2015localized} and \citet{LWC16} in recovering sparse eigenfunctions when the latent structure of the functional sample is sparse. Thus, we consider this as the state of the art method. The localized domains obtained by SFPCA do not perfectly equal zero and are therefore also hard-thresholded, this time with cut-off values $10^{-4}$ and $10^{-5}$ as these gave best performance results. Implementation details are provided in the supplementary material.
\end{enumerate}

All simulation results were obtained using the statistical software \texttt{R} version 4.4.1 \citep{RSoftware} on a PC running macOS Sonoma 14.4 with $18$ GB of RAM. Computational details, simulation study details, and code to replicate the results are available in the supplementary material.

\subsection{Recovering Underlying Supports}
\label{sec:LFPCA_sim}
We assess L-FPCA's ability to accurately identify localized eigenfunctions with supports $\mathcal{S}_1$, $\mathcal{S}_2$, and $\mathcal{S}_3$, for sample sizes $N \in \{250, 500, 1000\}$. 
We first smooth the noisy data \eqref{eq:noisy_data} to estimate the latent curves and capture genuine correlations while mitigating noise. Two main strategies exist: One fits smooth basis functions directly to the data \citep{ramsay2005functional}, and the other smooths the empirical covariance matrix to extract eigenfunctions and scores via spectral decomposition \citep{staniswalis98, Yao05, di09, goldsmith2013}. We adopt the second strategy using FPCA-FACE, reconstructing the smooth curves via KL decomposition with a number of principal components explaining 99\% of the variance of the sample. The high percentage of variance explained aims at reconstructing the primary features while discarding noise. Then, we identify the ``closest matching" true eigenfunction for each localized eigenfunction one using the Pearson correlation.

We compute specificity and precision to assess the performance of recovering the grid points outside the compact supports $\mathcal{S}_k$: 
\begin{align*}
 \text{Specificity}  = \frac{\text{TN}}{\text{TN} + \text{FP}} \,, \quad
    \text{Precision} = \frac{\text{TP}}{\text{TP} + \text{FP}} \,.
\end{align*}
%\lauramarginnote{the fact that we compute the metrics compared to the "best interval", using a tolerance, needs to be discussed\\
%{\color{blue}Above we mention that we match it using correlation. Isn't this what we do?}}
As the eigenfunctions equal zero outside the supports, this is identical to checking if we correctly estimate zero values of the eigenfunctions. Here, TN denotes the number of correctly identified grid points in the support, TP the number of correctly identified grid points outside the support, and FP the number of incorrectly identified grid points outside the support. Consequently, for the estimated compact supports $\widehat{\mathcal{S}}_k$ of $\mathcal{S}_k$, we have that ${\rm TN} = | \widehat{\mathcal{S}}_k^c \cap \mathcal{S}_k^c  |$, ${\rm TP} = | \widehat{\mathcal{S}}_k \cap \mathcal{S}_k |$, and ${\rm FP} = | \widehat{\mathcal{S}}_k  \cap \mathcal{S}_k^c |$, where $ | \mathcal{S}_k| = {\rm card}(\mathcal{S}_k) $ is the number of grid points contained in $\mathcal{S}_k$.

% Assessing the performance inside the compact supports cannot be done using sensitivity, as zero values in the eigenfunction can also occur in the compact supports. However, sensitivity is computed using false negatives, i.e., ${\rm FN} = | \widehat{\mathcal{S}}_k^c \cap \mathcal{S}_k |$, and it would thus decrease when the zero values of eigenfunctions within compact supports are correctly identified. Consequently, the results would be misspecified. We therefore compute the root mean squared error (RMSE) as 
% $${\rm RMSE}_{k,j} = \sqrt{\frac{1}{|\mathcal{S}_k|} \sum_{s \in \mathcal{S}_k} \left( \psi_{k,j}(s) - \widehat{\psi}_{k,j} (s)\right)^2}, \quad k=1,2,3, \, j=1,2,$$
% measuring the distance between the true eigenfunction $\psi_{k,j}(\cdot)$ and the estimated one $\widehat{\psi}_{k,j} (\cdot)$ over the corresponding $\mathcal{S}_k$.

\begin{figure}[htbp]
    \centering
\includegraphics[width=0.6\linewidth]{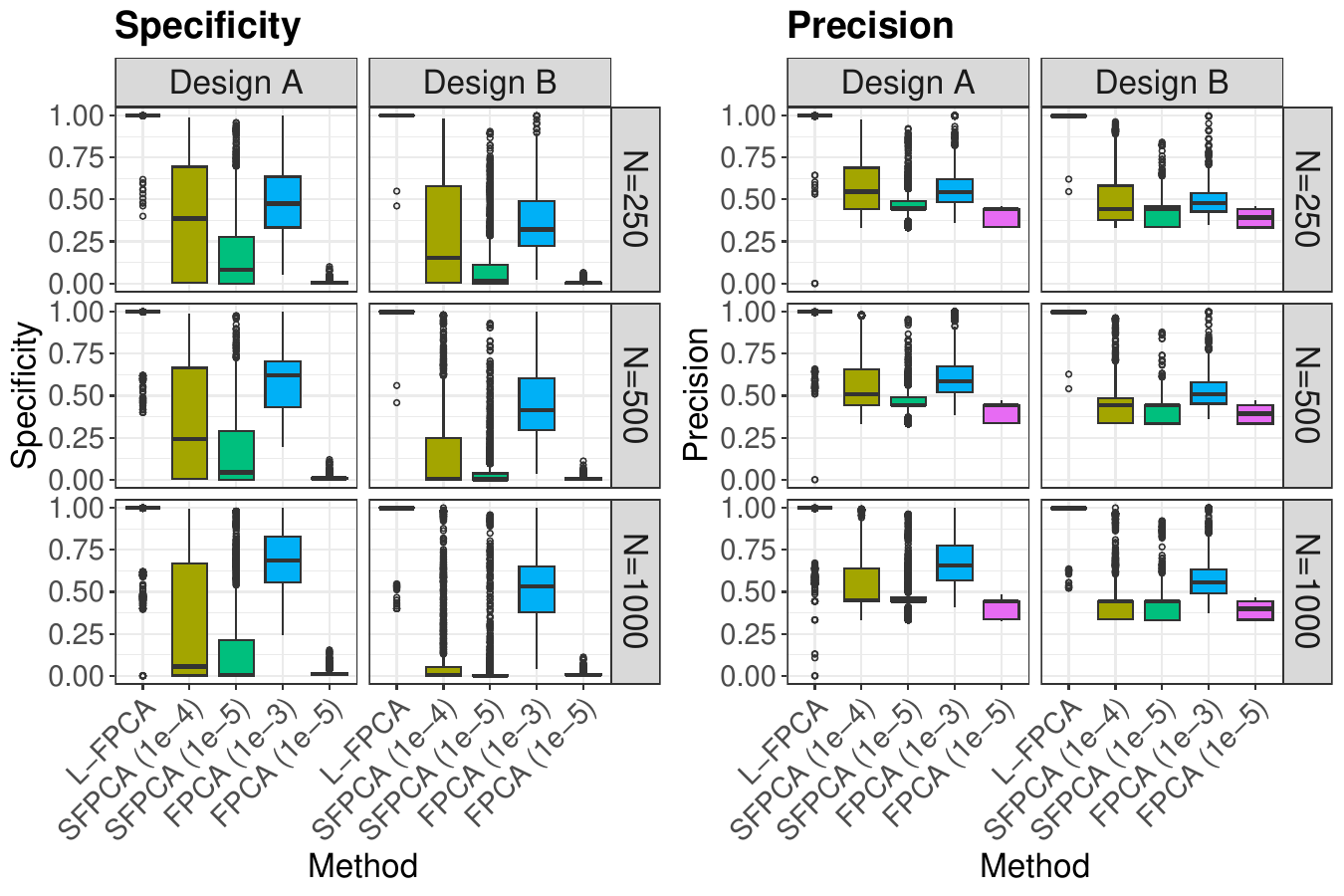}
    \caption{Visualization of specificity (\textit{left}) and precision (\textit{right}) for L-FPCA, SFPCA with thresholds $10^{-4}$ and $10^{-5}$, and FPCA($\tau$) with thresholds $10^{-3}$ and $10^{-4}$. For both design A (\textit{left column}) and design B (\textit{right column}), there are 500 realizations for sample sizes $N \in \{250, 500, 1000\}$.}
    \label{fig:boxplots_sim}
\end{figure}

Figure~\ref{fig:boxplots_sim} presents the results of our simulation study. L-FPCA achieves the highest specificity and precision in all the settings and across both designs.
%and exhibits the smallest RMSE for both simulation designs across all sample sizes for both designs. 
For the other two methods, specificity increases for a larger threshold, as more regions fall below this threshold in absolute terms. Precision seems comparable for SFPCA at both thresholds, while it decreases with $\tau$ for FPCA($\tau$).

%Regarding specificity, L-FPCA closely matches FPCA thresholded at 1e-5—both yield empirical distributions centered near one—while all other methods produce distributions centered at lower values. At the same time, L-FPCA achieves the highest precision across every scenario, whereas FPCA with a 1e-5 threshold performs worst in this regard. Finally, in both simulation designs and for all sample sizes, L-FPCA attains the lowest RMSE.

% \begin{figure}[htbp]
%     \centering
% \includegraphics[width=0.8\linewidth]{combined_plots_compact.pdf}
%     \caption{Visualization of specificity and precision (\textit{left}) and RMSE (\textit{right}) for L-FPCA, SFPCA with thresholds $10^{-4}$ and $10^{-5}$, and FPCA($\tau$) with thresholds $10^{-3}$ and $10^{-4}$. For both design A (\textit{top}) and design B (\textit{bottom}), there are 500 realizations for sample sizes $N \in \{250, 500, 1000\}$.}
%     \label{fig:boxplots_sim}
% \end{figure}

\subsection{Explained Variance}
\label{sec:subprocesses_sim}

A second motivation for L-FPCA is that the methodology preserves the explained variance of the principal components as the sparse structure of the underlying stochastic process is mirrored, without imposing overly localized eigenfunctions through regularization. In particular, as established in Lemma~\ref{l:EigenfunctionsAlign}, the explained variance for any sub-process $\mathcal{X}_{k'}(\cdot)$ is given as
 \begin{equation*}
    \label{eq:pve_subprocesses}
        \frac{ \sum_{j=1}^{J_{k'}} \lambda_{k', j} }{ \text{totVar}}  \,, \quad  \text{totVar} = \sum_{l=1}^{L} \lambda_{l} = \sum_{k=1}^K \sum_{j=1}^{J_{k}} \lambda_{k, j} \,.
    \end{equation*}
In practice, the eigenvalues are approximated by their sample counterparts $ \widehat{\lambda}_{k,j}$ and a bound for the approximation error is given in Remark~A1 in the supplementary material.

In this section, we assess the preservation of variance explained by computing the percentage of variance explained (PVE) in \eqref{eq:pve_subprocesses}. Specifically, we examine how the estimated PVE for each sub-process $\mathcal{X}_k(\cdot)$ on $\mathcal{T}_k$ differs from the true PVE on $\mathcal{T}_k$ for all methods discussed in Section~\ref{sec:LFPCA_sim}. For design A and B, the true PVE for $\mathcal{X}_1(\cdot)$ is $\text{PVE}_1 = \lambda_{1,1}/\text{totVar}$ and $\text{PVE}_1 = (\lambda_{1,1} + \lambda_{1,2})/\text{totVar}$ respectively. The true PVE for the remaining two sub-processes is given as $\text{PVE}_k = \lambda_{k,1}/\text{totVar}$ for $k\in\{2,3\}$ across both simulation designs. The estimated PVE is obtained by replacing the population eigenvalues by their sample counterparts.

\begin{figure}[htbp]
    \centering
    \includegraphics[width=0.3\linewidth]{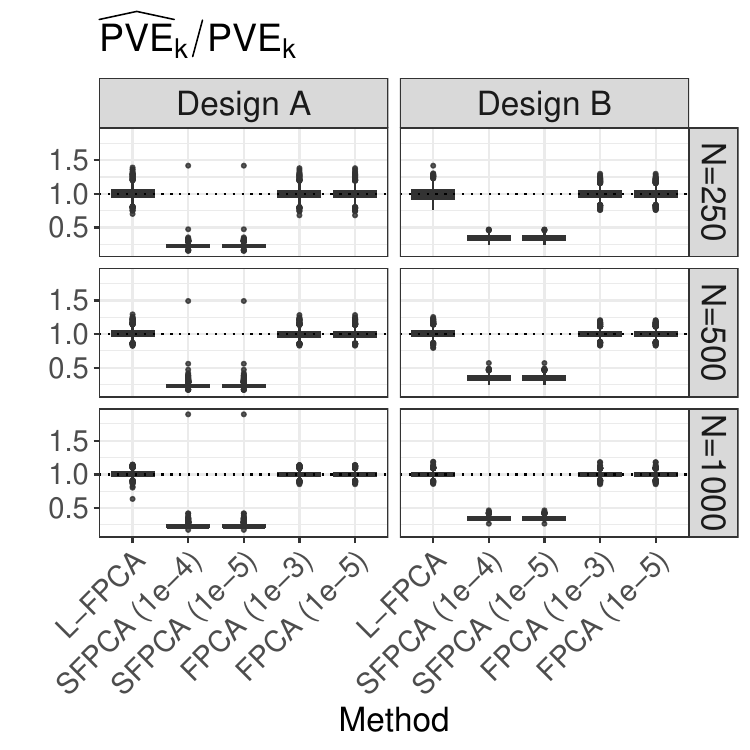}
    \caption{Ratio $\widehat{\text{PVE}_k}/\text{PVE}_k$ of estimated and true PVE for L-FPCA, SFPCA with thresholds $10^{-4}$ and $10^{-5}$, and FPCA($\tau$) with thresholds $10^{-3}$ and $10^{-4}$. The ratio is summarized in a single box-plot summarizing all three sub-processes $k\in\{1,2,3\}$ across simulation designs A (\textit{left)} and B (\textit{right}) for $N \in\{250,500,1000\}$. }
    \label{fig:boxplots_subprocesses_sim}
\end{figure}

In Figure~\ref{fig:boxplots_subprocesses_sim}, we visualize the ratio $\widehat{\text{PVE}_k}/\text{PVE}_k$, $k=1,2,3$, for all sub-processes across both simulation designs with sample size $N \in\{250,500,1000\}$. L-FPCA performs comparably to thresholded FPCA across all scenarios, while SFPCA systematically underestimates the true PVE.

\section{Real Data Examples}
\label{sec:RealData}

\subsection{Pinch Force Curves}
We start by analyzing data containing $N=20$ force‑meter traces of brief thumb–forefinger pinches: Each trial begins with the subject holding a set baseline force, squeezing to a specified peak, then returning to that baseline. The time domain $\mathcal{T}= [0, 0.30]$ spans over $0.3$ seconds (300 milliseconds) on a regular grid with $P=151$ points. The data set is available in the \texttt{fda} package \citep{fda}, and additional information are available in \citet{ramsay2005functional}. 

\begin{figure}[htbp]
    \centering
    \includegraphics[width=0.8\linewidth]{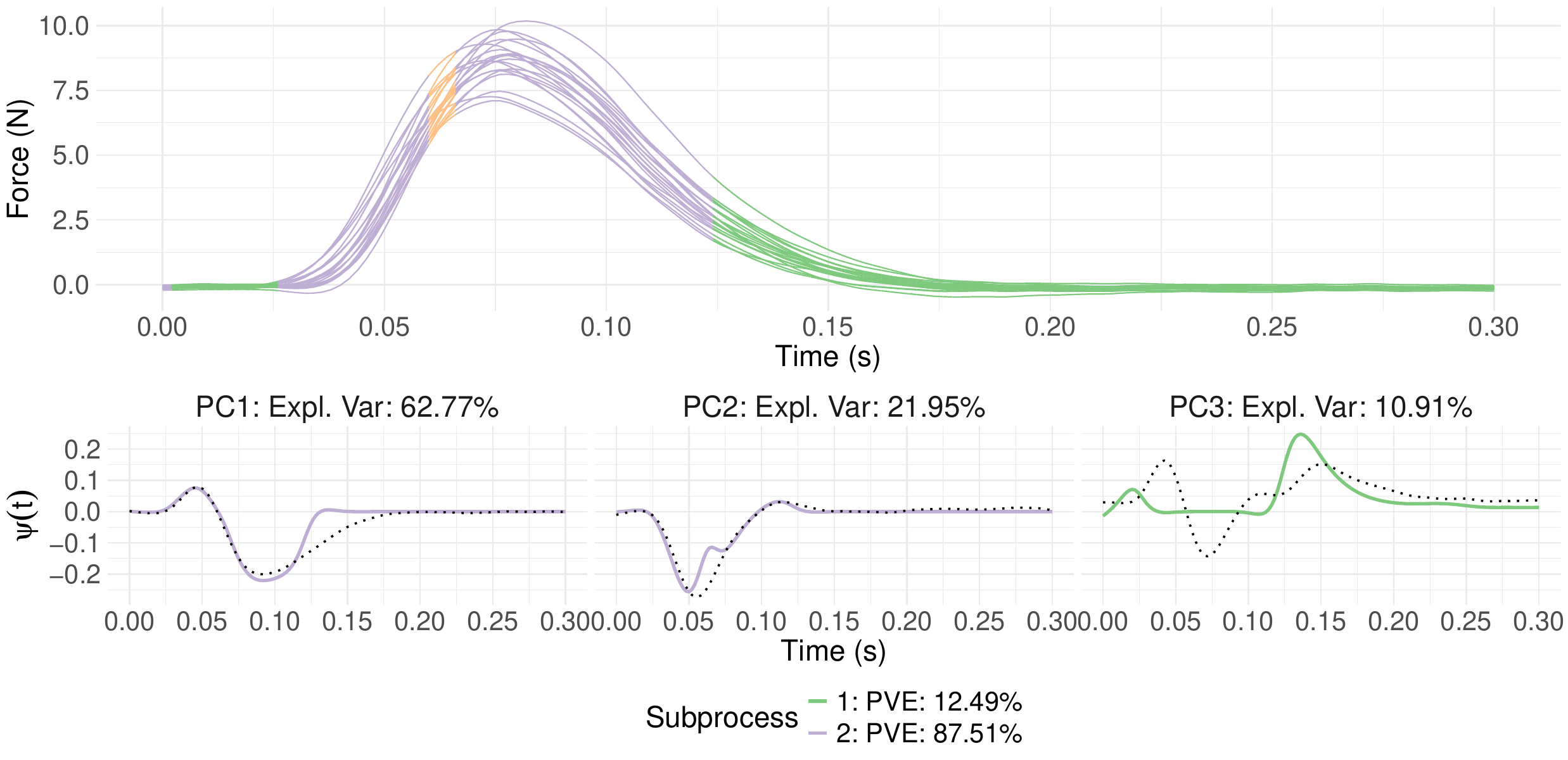}
    \caption{
    \textbf{Top:} Smoothed pinch force curves, with colored segments highlighting the disjoint sub-intervals of the uncorrelated sub-processes identified by our methodology. \\
    \textbf{Bottom:} First three localized eigenfunctions obtained using L-FPCA, colored according to the their support. The PVE of each eigenfunction is shown above each plot, and the PVE of each sub-process is provided in the legend. For comparison, the eigenfunctions obtained from FPCA are visualized as dashed curves (\textit{dotted black}).
    }
    \label{fig:pinch}
\end{figure}

In Figure~\ref{fig:pinch}, we visualize the data curves. As in Section~\ref{sec:SimulationStudy}, we smoothed curves using a truncated KL representation via FPCA-FACE with PVE as high as 99\%, and we estimated $M=6$ localized eigenfunctions. The curves are color coded according to the sub-processes detected by BD-SVD. Three disjoint supports are detected: The time before and after the pinch (\textit{green}), the pinch itself (\textit{violet}), and a short moment before the peak of the pinch (\textit{orange}). Further, we show the first three computed localized eigenfunctions.

The first localized eigenfunction accounts for over 62\% of the sample variability and reveals a discordant pattern: It represents stronger-than-average force before the peak, and weaker-than-average force afterwards, or vice versa. Moreover, the sub-process corresponding to the moment of the pinch (\textit{violet}) explains $87.51 \%$ of the sample variability, and thus most of it. This is plausible, as subjects seem to exhibit the greatest behavioral differences during this phase of the experiment (Figure~\ref{fig:pinch}).

The third localized eigenfunction captures variability both before and after the pinch, with greater weight on the post-pinch period. This is consistent with the functional data, which show more variation after the pinch than before. In contrast, the third eigenfunction obtained via FPCA (\textit{dotted black}) does not isolate variability specifically around the pinch process.

We note that the visualized localized eigenfunctions that belong to two (\textit{green} and \textit{violet}) out of all three sub-processes explain more than $99\%$ of the observed variability. Consequently, the remaining sub-process (\textit{orange}), and therefore the moment before the peak of the pinch, contributes very little to the explained variability in the experiment.

\subsection{EEG Curves of Control and Alcoholic Groups}
Next, we apply L-FPCA to electroencephalography (EEG) records of alcoholic patients compared to a control group. For these groups, the recordings contain $N_A =77$ and $N_C =45$ signals, respectively, from $64$ electrodes placed on different locations of the heads of the patients which are recorded at 256 Hz (3.9-ms epoch) in one second. Thus, the EEG curves have domain $\mathcal{T} = [0, 256]$ which is regularly and densely discretized over a grid with $P=256$ points. Data is publicly available at \url{https://kdd.ics.uci.edu/databases/eeg/eeg.html} or in the supplementary material of \citet{xue2024optimal}. As in the previous application, EEG curves were pre‑smoothed by a truncated KL expansion via FPCA‑FACE, capturing 99\% of the variance.

\begin{figure}[htbp]
    \centering
    \includegraphics[width=0.8\linewidth]{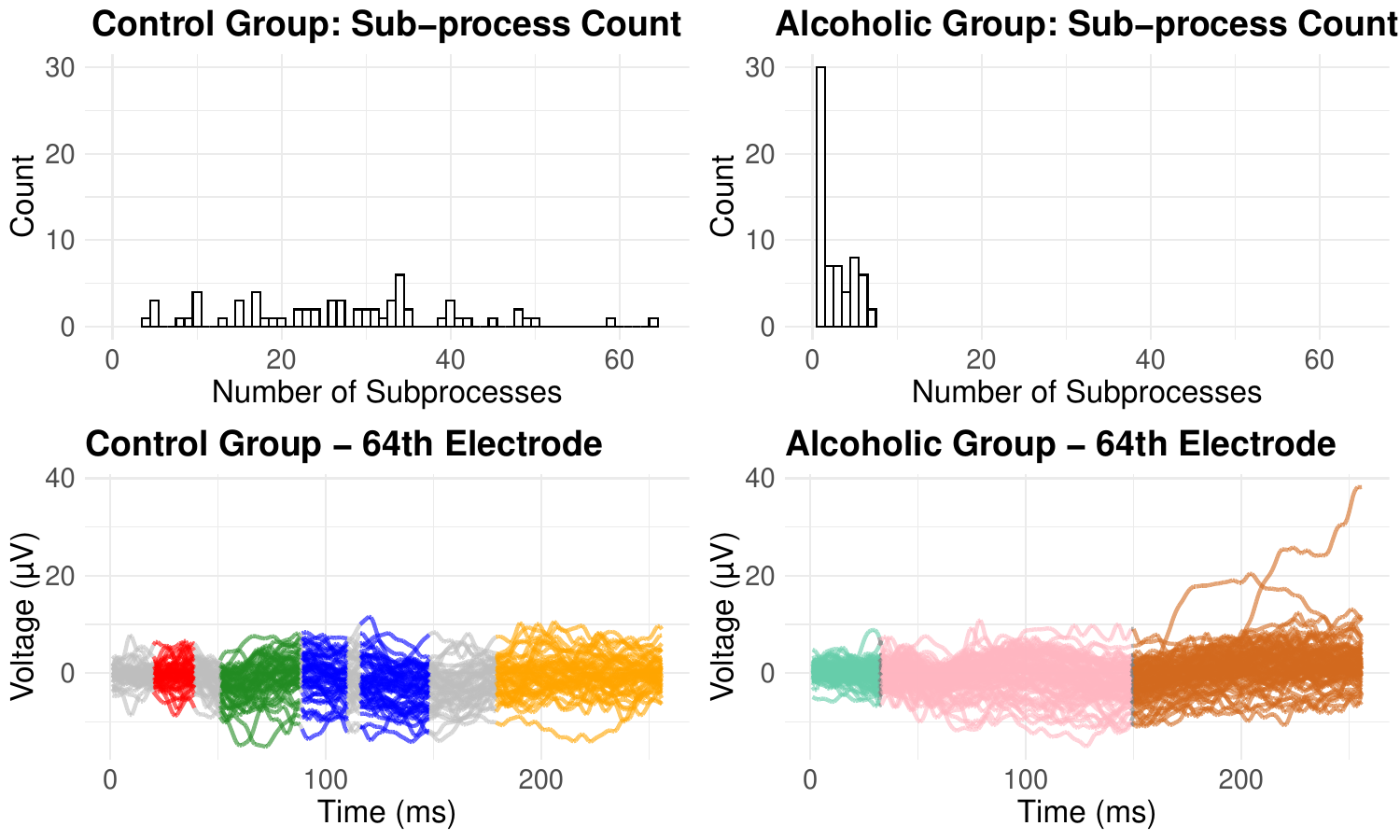}
    \caption{
    \textbf{Top:} Number of identified disjoint sub-intervals of the uncorrelated sub-processes across all electrodes for the control group (\textit{left}) and alcoholic group (\textit{right}). \\
    \textbf{Bottom:} Smoothed EEG signals from the 64th electrode for the control (\textit{left}) and alcoholic group (\textit{right}). The colored segments highlight the disjoint supports of the sub-processes exhibiting the first six localized eigenfunctions. Gray segments indicate the remaining intervals, where larger gray areas correspond to multiple intervals grouped together.}
    \label{fig:hist_and_curves}
\end{figure}

The number of distinct sub-processes detected by L-FPCA differs between the alcoholic and control groups: alcoholic patients exhibit fewer sub-processes, indicating that their EEG curves remain autocorrelated over longer intervals. This contrast is visualized in the top row of Figure~\ref{fig:hist_and_curves}, which displays the number of detected sub-processes across all electrodes.

The bottom row of Figure~\ref{fig:hist_and_curves} shows the smoothed EEG signals from electrode 64, colored according to the identified disjoint supports of the uncorrelated sub-processes that exhibit the first six localized eigenfunctions. Other sub-intervals are shown in gray scale, and larger gray areas correspond to multiple intervals grouped together.  The control group decomposes into 23 sub-processes, whereas the alcoholic group comprises only three blocks.

\begin{figure}[htbp]
    \centering
    \includegraphics[width=0.7\linewidth]{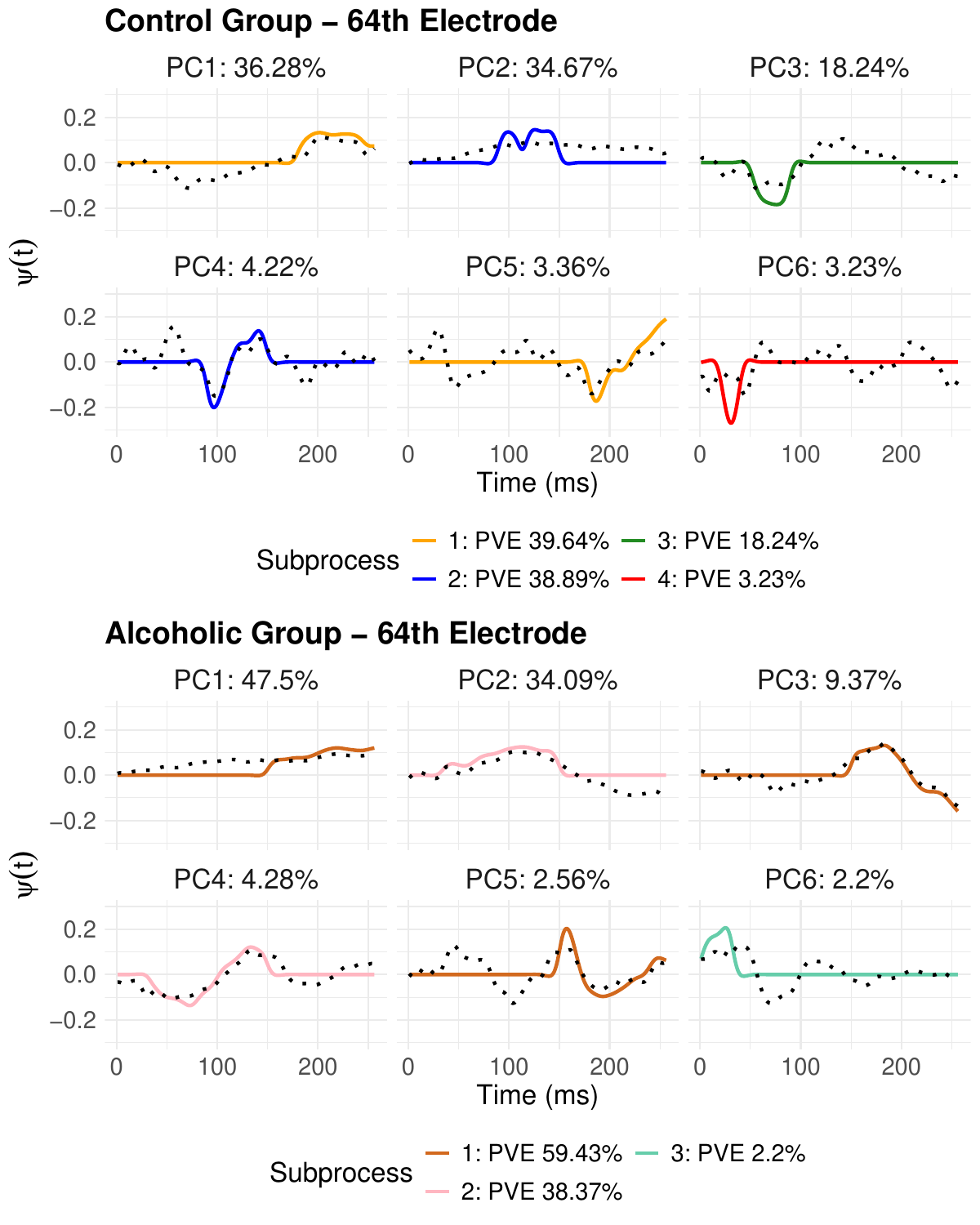}
    \caption{
    First six localized eigenfunctions obtained via L-FPCA for the EEG curves recorded from the 64th electrode for control group (\textit{top}) and alcoholics (\textit{bottom}). Each plot provides the variance explained by each component, and different colors highlight disjoint sub-intervals  corresponding to distinct sub-processes. Their overall PVE is given in the legend. For comparison, the eigenfunctions obtained from FPCA are also displayed (\textit{dotted black}).
    }
    \label{fig:loc_eigenfunctions}
\end{figure}

Figure~\ref{fig:loc_eigenfunctions} visualizes the corresponding first six localized eigenfunctions. For comparison, we also include the estimated eigenfunctions obtained from FPCA without localization (\textit{dotted black}). The first three plots of eigenfunctions of the alcoholic group are identical to the illustrative example in Figure~\ref{fig:Introduction} in the introduction of this work. As discussed above, the eigenfunctions for the control and alcoholic group are based on four and three disjoint supports, respectively.

%In other words, the control group’s main trend involves remaining either consistently above or below the sample mean over specific sub-intervals. In the alcoholic group, however, a range of more complex deviations from the sample mean is apparent, reflecting a richer variety of underlying functional dynamics. 

The localized eigenfunctions indicate that, for both the control and alcoholic groups, the majority of the variance is concentrated in the middle and towards the end of the recording period, while eigenfunctions supported near the initial time segments account for comparatively little variance. This observation is underpinned by the PVE for the identified sub-processes. In the control group, two sub-processes (\textit{blue} and \textit{orange}) exhibit similar PVEs, meaning that they contribute equally to the total variability (Figure~\ref{fig:hist_and_curves} and Figure~\ref{fig:loc_eigenfunctions}). Together, these sub-processes account for almost 80\% of the total variance explained. In contrast, the alcoholic group shows a more concentrated variance structure, with a single sub-process (\textit{brown}) contributing nearly 60\% to the total variance, suggesting that the dominant EEG patterns emerge primarily during the final third of the recording window (Figure~\ref{fig:hist_and_curves} and Figure~\ref{fig:loc_eigenfunctions}). These findings enable a more targeted and parsimonious strategy to medical signal analysis: rather than processing the entire signal uniformly, clinicians can focus on those sub-intervals corresponding to the dominant latent sub-processes.

We provide plots similar to Figure~\ref{fig:hist_and_curves} for curves of three other electrodes (for the 16th, 32nd, and 48th electrode) in the supplementary material.

% This suggests that the alcoholic group’s underlying functional dynamics are fundamentally different, reflecting a richer variety of deviations from the sample mean than those seen in the control group.

\section{Discussion}
\label{sec:Discussion}
% In this work we suggested a method to perform localized functional principal component analysis. As a natural extension of this, we proposed a way to uncover compactly supported functional sub-processes of the overall process a functional sample is a realization of. The key to these relies on the detection of the block structure of the empirical covariance function. Specifically, if the functional observations are realization of a continuous process characterized by a strong autocorrelation in parts of its domain, then sub-intervals of it are detected and used as supports of the localized eigenfunctions. Then, with L-FPCA we are not only able to understand which are the main directions of variation of the overall sample, but also to quantify how much every sub-process contributes to the overall variation of the sample. Thus, our method proves to be crucial when understanding the amount of variation explained by specific parts of the domain, as shown by our application to EEG data.
In this work, we introduce a novel localized functional principal component analysis (L-FPCA) method that not only captures the main modes of variation in functional data but also quantifies the contribution of distinct sub-processes. By detecting block structures in the empirical covariance function, our approach identifies sub-intervals with strong autocorrelation and uses them as supports for compactly supported eigenfunctions. Because these localized sub-processes sum to the overall process, our method allows us to determine the proportion of total variation explained by each individual sub-process. This feature is particularly advantageous when analyzing continuous data observed over large, dense grids, as it enables a parsimonious focus on subsets of the domain that account for a substantial share of the overall variation. Overall, L-FPCA offers an interpretable and efficient framework for decomposing complex functional data, as demonstrated in our application to EEG recordings.

Several promising avenues exist for extending our work. First, our method naturally extends to functional observations over $\mathbb{R}^d$ for $d > 1$, which broadens its applicability to multidimensional domains \citep{shamshoian2022bayesian, wang2022low}. In addition, applying our approach to stochastic processes with well-studied autocorrelation structures, such as time series and random fields, could provide valuable insights. Extending the method to multivariate functional data, where the covariance structure is inherently more complex than in the univariate setting \citep{li2020fast}, also represents an important direction for future research. Finally, the ability to identify localized sub-processes opens the door to classification applications. Clustering methods for functional data have become increasingly popular, see for instance \citet{james2003clustering, tarpey2003clustering, ieva2016covariance, delaigle2019clustering,  zeng2019simultaneous, wang2024multiclass}.

\section*{Supplementary Material}

\begin{description}
\item[Online Appendix:] This online appendix contains the derivations and proofs of the results presented in this paper, a detailed recap of BD-SVD, computational details for the simulation studies and real data examples, and additionally supplementary material for the EEG recordings example.
\item[Code:] \texttt{R} code for the implementation of L-FPCA and for the replication of the examples in Section~\ref{sec:RealData} and the simulation studies in Section~\ref{sec:SimulationStudy}.
\end{description}

All supplementary files are provided in a single archive.

\section*{Disclosure Statement}
The authors have no conflicts of interest or competing interests to declare.

\bibliographystyle{chicago}

\bibliography{Bibliography}
\end{document}